# DesignCon 2011

# Comparison of Optical and Electrical Links for Highly-Interconnected Systems

James Kruchowski, Mayo Clinic

Vladimir Sokolov, Mayo Clinic

Nathan E. Harff, Mayo Clinic

Mark A. Nelson, Mayo Clinic

KY Liou, Multiplex, Inc.

Graham Cameron, Mayo Clinic

Barry K. Gilbert, Mayo Clinic
gilbert.barry@mayo.edu, 507-284-4056

Erik S. Daniel, Mayo Clinic

## Abstract
As data rates for multi-gigabit serial interfaces within multi-node compute systems approach and exceed 10 Gigabits per second (Gbps), board-to-board and chip-to-chip optical signaling solutions become more attractive, particularly for longer (e.g. 50-100 cm) links.  The transition to optical signaling will potentially allow new high performance compute (HPC) system architectures that benefit from characteristics unique to optical links.

To examine these characteristics, we built and tested several optical demonstration vehicles; one based on dense wavelength division multiplexing (DWDM), and others based on multiple point-to-point links carried across multimode fibers.  All test vehicles were constructed to evaluate applicability to a multi-node compute system.  Test results, combined with data from recent research efforts are summarized and compared to equivalent electrical links and the advantages and design characteristics unique to optical signaling are identified.

## Authors Biographies
James Kruchowski is a Senior Project Engineer for the Special Purpose Processor Development Group at Mayo Clinic.  He received his B.S. degree in Electrical Engineering, with distinction, in 1980 from the University of Minnesota.  He joined Mayo in 2001, after having previously served as a signal integrity team lead and PHY characterization group manager for Cray Research and Intel.  He is currently involved with research programs in high performance optics, link characterization, systems engineering and signal integrity.

Vladimir Sokolov is a Lead Engineer for the Special Purpose Processor Development Group at Mayo Clinic.  He received his B.S. in Science Engineering from Northwestern University and the M.S. and Ph.D. in electrical engineering in 1973 from the University of Wisconsin, Madison.  At Mayo since 2001, his research interests continue to be in the areas of microwave and mm-wave integrated circuits and more recently in optical interconnects for high speed data transfer for HPC applications.

Nathan Harff is a Senior Project Engineer for the Advanced Technology Section of the Special Purpose Processor Development Group at Mayo Foundation.  He received his B.S. in Engineering Physics from North Dakota State University and his Ph.D. in electrical engineering in 1997 from Oregon State University specializing in solid-state physics.   He joined Mayo in 2001, after holding post-doctoral fellow positions at the Max Planck Institute for Solid-State Physics and at Sandia National Laboratories.  His research interests include optical interconnects for high speed data transfer and thermal management for HPC applications.

Mark Nelson is an Engineer in the Lab and Technologies Section of the Special Purpose Processor Development Group at Mayo Foundation.  He received his B.S. in Electrical Engineering  from the University of Minnesota. At Mayo since 1999, his research interests continue to be in the areas of optics, testing, programming and recently in optical interconnects for high speed data transfer for HPC applications.

K.Y. Liou is Director of Technology and Government Business at Multiplex, Inc. Prior to joining Multiplex, he had 20 years of experience with Bell Labs in Holmdel, New Jersey.  He has made numerous contributions to the research, development, and manufacturing of single-frequency lasers, photonic integrated circuits, opto-electronic integrated devices, and WDM long-haul, metro and access systems. At Multiplex, Inc., he has managed R&D programs funded by U.S. government, including DARPA, Air Force Research Lab, and Army Research Lab programs, as well as new technology and product development for optical communication networks. He has a B.S. degree from National Taiwan University and Ph.D. from University of Wisconsin – Madison.

Graham P. Cameron is a Senior Project Engineer for the Special Purpose Processor Development Group at Mayo Clinic. He received the B.A. Degree in Economics from Carleton College, and B.S., M.S., and Ph.D. degrees in Electrical Engineering from the University of Minnesota with a concentration in magnetics. His present research interests include the electron-spin dependent devices, quantum computing, signal integrity, and medical applications of magnetism. His previous work includes magnetic recording head design and tester development at Seagate Technology Inc.; piezoelectric shock sensor and low-power digital product development at Event Tracking Service; electron-spin probing of magnetic surfaces at Hitachi's Advanced Research Laboratory; and magnetic transducer and magnetic random access memory development at Honeywell.

Barry K. Gilbert received a BSEE from Purdue University (West Lafayette, IN) and a Ph.D. degree in physiology and biophysics from the University of Minnesota (Minneapolis, MN). He is currently Director of the Special Purpose Processor Development Group at the Mayo Clinic, directing research efforts in high performance electronics and related areas.

Erik S. Daniel received a BA in physics and mathematics from Rice University (Houston, TX) in 1992 and a Ph.D. degree in solid state physics from the California Institute of Technology (Pasadena, CA) in 1997. He is currently Deputy Director of the Special Purpose Processor Development Group at the Mayo Clinic, directing research efforts in high performance electronics and related areas.

# Introduction

Current multi-gigabit serial interface designs used within HPCs incorporate large numbers of node-to-node digital links with data rates that extend to 10 Gbps and beyond. Some of the system designs have extended board-to-board signal reach requirements. Electrical signaling has historically been used in these systems for shorter intra-system links, and active cables (both optical and electrical) are beginning to gain market acceptance for longer links (~ > 2 m). Building faster links based on electrical signaling requires progressively more sophisticated compensation techniques, such as signal equalization and pre-emphasis, to maintain desired link bit error rates as the respective data rates and link lengths continue to increase. Electrical signaling appears to be nearing maximum potential data rates for these link configurations and lengths due to the high power required for signal compensation at higher data rates; hence it is anticipated that optical signaling may soon be the "wave of the future" for these serial links within HPC [1]. Indeed, optical interconnects have been suggested for over 25 years as the "wave of the future" [4], and yet have not gained universal acceptance for HPC design applications. To gain some perspective on this question, we sought to better understand the mechanics of the transition from electrical to optical in HPCs, by surveying the status of current optical research and actually assembling scaled-down versions of candidate interconnect systems using state-of-the-art components.

Several major research laboratories are conducting research aimed at developing integrated optical technology specifically for the shorter reach applications (intra-board and even intra chip) anticipated in future HPC systems [1][2][3]. These efforts are aimed primarily at miniaturizing optical components and reducing their power consumption so that they can be integrated cost effectively on silicon. At the system level, HPC application opportunities were suggested from previous programs, including the Multiwavelength Assemblies for Ubiquitous Interconnects (MAUI) and Chip-to-Chip Optical Interconnect (C2OI) programs sponsored by DARPA [5][9]. While active optical cables (AOCs) appear as an emergent application of optics to HPCs [6], it is generally acknowledged that multimode optical components should begin to appear more commonly in HPC designs within the next 1-3 years, while single-mode optics applications are at best, five to ten years away from being implemented in actual HPC system designs. The hardware demonstrations described in this paper specifically utilize available technology and infrastructure to demonstrate and compare two fundamentally different approaches including a broadcast and select system utilizing DWDM, and a "parallel optics" system that essentially takes advantage of spatial multiplexing.

The efforts described within this paper extended the previous efforts to include the design and construction of a series of optical test vehicles to examine their respective applicability to HPC designs. At the start of this paper, a brief discussion of the HPC system application background is provided, along with the primary objectives of the optical demonstration test vehicles. Three distinct optical link test vehicles are then described which were constructed to evaluate potential system benefits and the relative technology maturity of each solution, in conjunction with evaluating the scalability associated with these link concepts. Key evaluation metrics are described and used throughout this document. The design characteristics of the optical test vehicles are described in detail, and the test results from each vehicle are summarized and discussed. Based upon these efforts the relevant electrical and optical signaling characteristics are compared and summarized along with the resultant conclusions.

## Highly Interconnected HPC System Designs

The primary compute system design used as a model for our scaled link comparison studies begins as a very large number of compute nodes, typically including local memory storage at each node. A conceptual example of this type of system design is shown below in Figure 1.

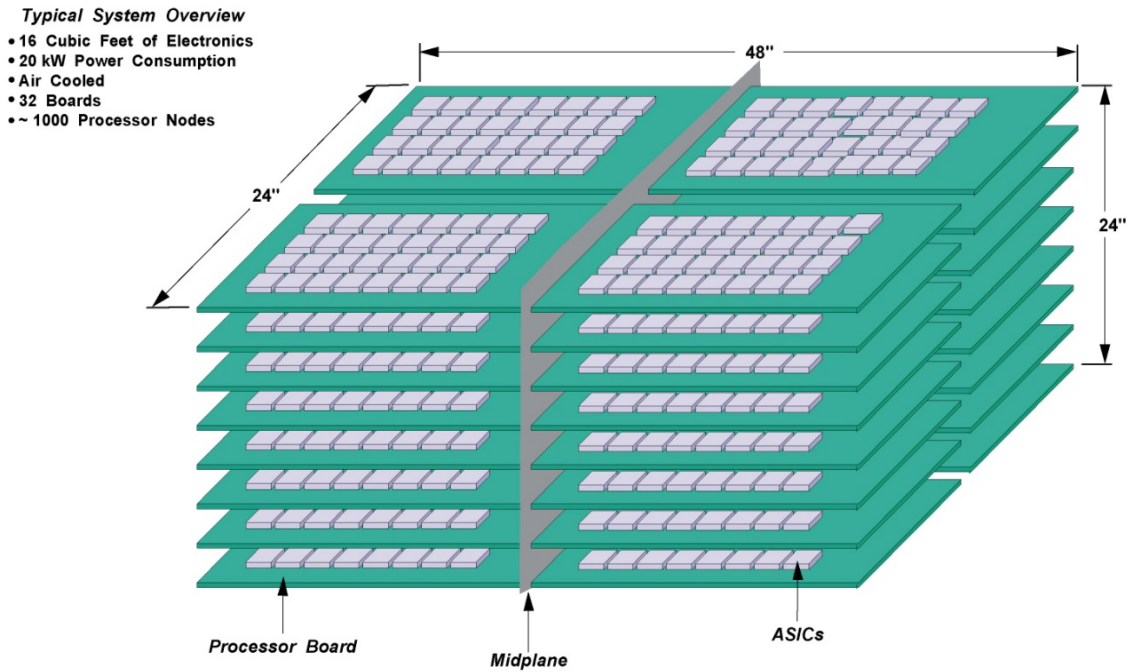

**Figure 1: Conceptual Example of a High Performance Compute (HPC) System (42075)**

One optimized interconnect solution for this class of computer design consists of an "all-to-all" communication network that links each and every processor node to all of the other processing nodes within the system. System designs in this class of machine can contain large numbers (>10,000) of high-speed communication links operating at data rates up to 10 Gbps. Lengths of these high-speed electrical transmission links may total 100 cm (approx. 40 in) in PCB tracelength, and may also include multiple parasitic effects such as those associated with board-to-board connectors, via transitions within a board, and device packaging.

In an all-to-all interconnection network a total of $N*(N-1)$ links are required for an N x N array of processing nodes. As will be discussed later, it seems optical components may provide interesting alternative methods for accomplishing this same goal. For example, the use of a star coupler to broadcast signals to all nodes is the basis for one of our test vehicles.

To evaluate the distribution of link lengths expected within a nominal all-to-all system, we computed the distribution of link lengths (based on Manhattan routing distance) needed to support a 6 x 6 array of processing nodes placed on a 10 cm pitch. The range of calculated link lengths is shown below in Figure 2:

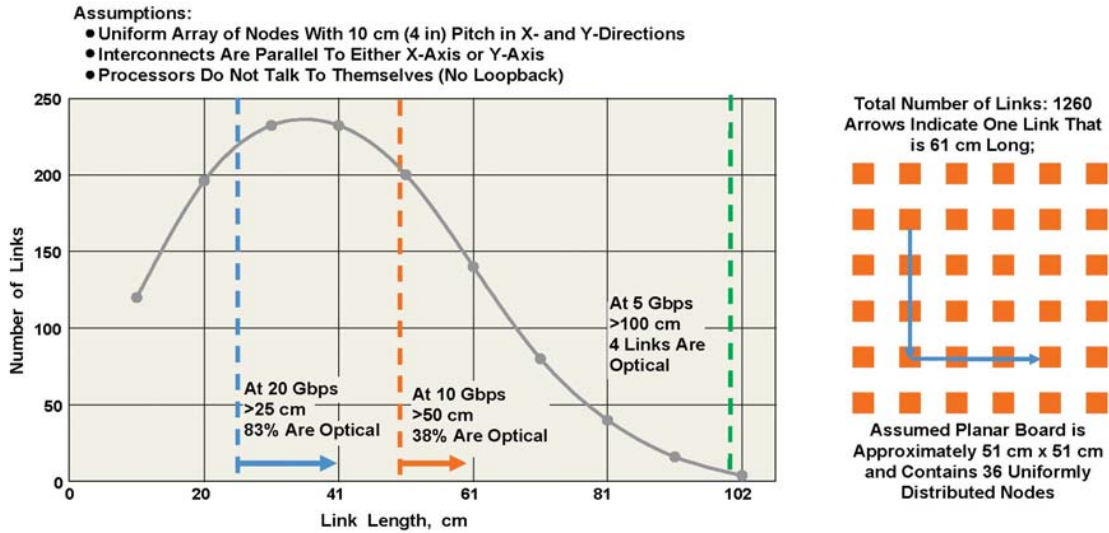

**Figure 2: Interconnect Link Length Model Based on a 6X6 Array of Processing Nodes Assuming an Electrical Bandwidth-Distance Product Limitation of 500 Gbps-cm (41424)**

The graph also shows the breakpoint and percentage distribution for electrical versus optical links within the system, based upon the bandwidth-distance (B*d) products of the links, assuming that electrical links are capable of only supporting up to a 500 Gbps-cm B*d product (roughly consistent with the current state of the art, see Figure 18 near the end of this document.) The B*d metric is explained in much greater detail in the following "Key Metrics" section.

As can be seen from this simplified interconnect model, with 5 Gbps serial links, only 4/1,260 links would necessitate the use of optical interconnects. At 10 Gbps nearly 38% of the link lengths would require the use of optical links, and at 20 Gbps, nearly 83% of the link lengths would require the use of optical links.

# Key Metrics Used For Evaluating Electrical and Optical Links

We identified four key metrics for consideration when evaluating electrical and optical links for a given HPC application. These are described in detail in this section.

## Bandwidth-Distance Product (B*d)

In the high-speed, low-power optical interconnect research community, a commonly used metric is the bandwidth-distance product (i.e. the product of the signal bandwidth and the communication link distance, B*d)[15]. In multimode fiber, for example, the B*d product is a constant, since the bandwidth limitation is primarily due to modal dispersion that is proportional to the length of the fiber. The B*d product is not necessarily a constant for electrical interconnects or even single-mode fiber. Nevertheless, for short links B*d is often cited to first approximation as a constant to facilitate the comparison of different types of links including electrical links. This paper follows the same use of the B*d metric to compare waveguides in a uniform fashion across all of the links considered.

Based upon our earlier literature search results, we found that electrical interconnect link performance is typically within the range of 250 to 1500 Gpbs-cm, with 500 Gbps-cm being the most commonly seen value [10]. Recent optical channels based upon a "hybrid" PCB construct with polymer-based optical channels and conventional electrical channels have shown the B*d for the polymer-based waveguides to be in the range of 1500-3000 Gbps-cm for measured values, with ideal limits up to 6000 Gbps-cm [9]. Further evaluation of multimode fiber optics shows B*d to range significantly higher, 4000 up to 500,000+ Gbps-cm. B*d for single-mode fiber extends beyond the range of multimode fibers (from 500,000 Gbps-cm and up). For HPC applications reaching the limits of electrical interconnect bandwidth and lengths, optical interconnects have the performance advantage for this metric.

## Power Per Link and Energy Per Bit Transferred

Another commonly used metric for high performance links within HPC system designs, where the number of links may extend into the 10s of thousands range and beyond, is the normalized power consumption (in mW/Gbps), that includes the power consumption of both the high-speed transmitter (Tx) and receiver (Rx). Typically the normalized link power for many of the links under consideration is in the milliWatt range. Equivalently, the normalized power can be expressed in energy per bit transferred, in terms of picoJoules/bit (pJ/b), where 1 pJ/b=1mW per Gbps.

## Cost

When developing computing systems with large numbers of processors and other components, cost is an important factor. In our research-based environment where we typically work within the boundaries of very small numbers of prototype test vehicles, it is difficult to establish valid cost numbers that accurately reflect high volume costs. However we were able to establish some degree of low-volume cost metrics during this project. Current cost for HPC DWDM links implemented in single-mode optics were found to be in the range of 10-50 $/Gb/s [8]. The cost of multimode links with VCSEL sources is significantly lower. Low volume costs for optical transceivers in the range of 5-16 $/Gb/s for a 10 Gb/s multimode transceiver, with high volume costs projected to be 1-5 $/Gb/s. These costs, especially for multimode optical components appear to be dropping rapidly over time. Discussions with research and design teams from other major HPC system development organizations have suggested that a crossover from electrical to multimode optical links may occur when the cost range of 1-2 $/Gb/s is reached. This metric is illustrated graphically below in Figure 3 [13].

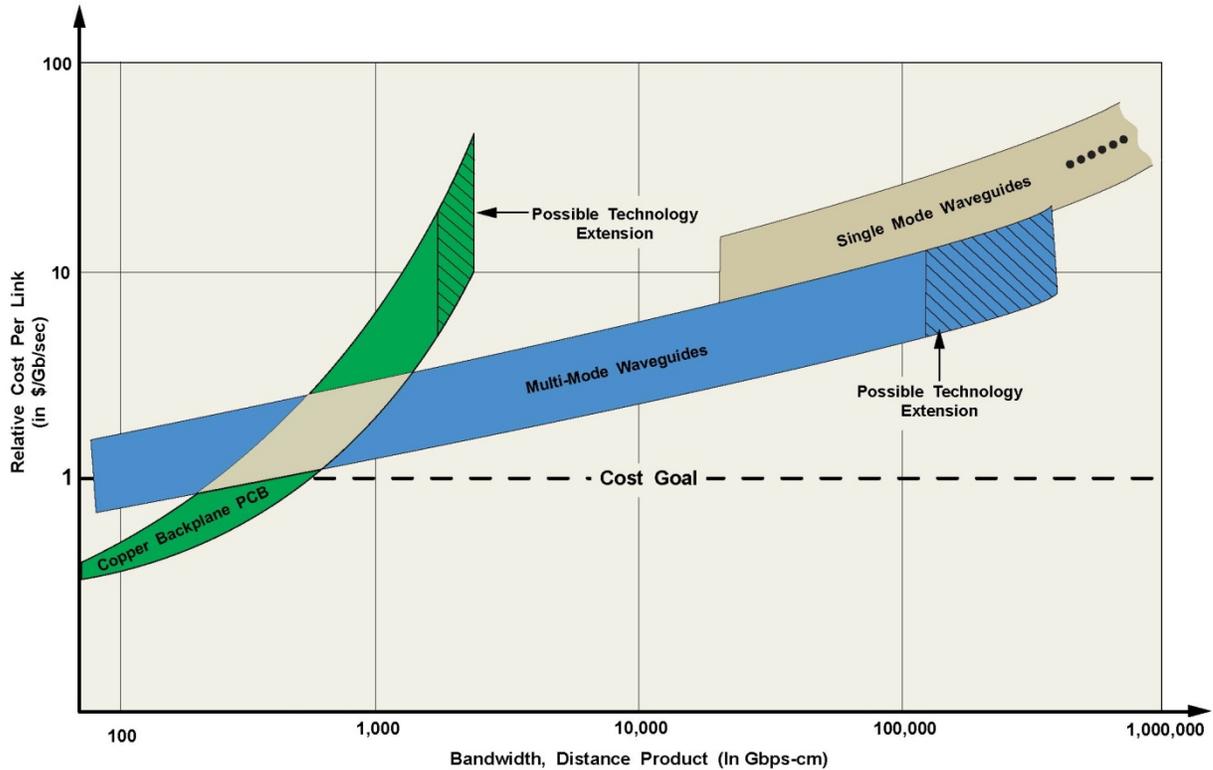

**Figure 3: Estimated "Crossover Zone" for Optical vs. Electrical Links Based Upon Cost and B*d (41428)**

At the present time, high-speed electrical links for short reach applications are still the more cost-effective solution for modest bandwidth-distance product applications when compared to their equivalent optical link counterparts.

## Interconnect Packaging Density

Within HPC and data center designs, the application of Active Optical Cables (AOCs) is becoming more prevalent [6]. We believe the primary motivations associated with this trend are driven by B*d, Power, Cost, and a fourth metric that we would refer to as spatial interconnect density, which may be expressed in terms of the interconnect bandwidth divided by the face surface area of the interconnect, or Gbps per mm$^2$. Two examples comparing achievable densities for multimode optical and electrical interconnect designs are illustrated in Figure 4. Although the optical packaging shown leaves a significant amount of unutilized open area, the interconnect density ratio still favors the optical interconnect by approximately 7 to 1.

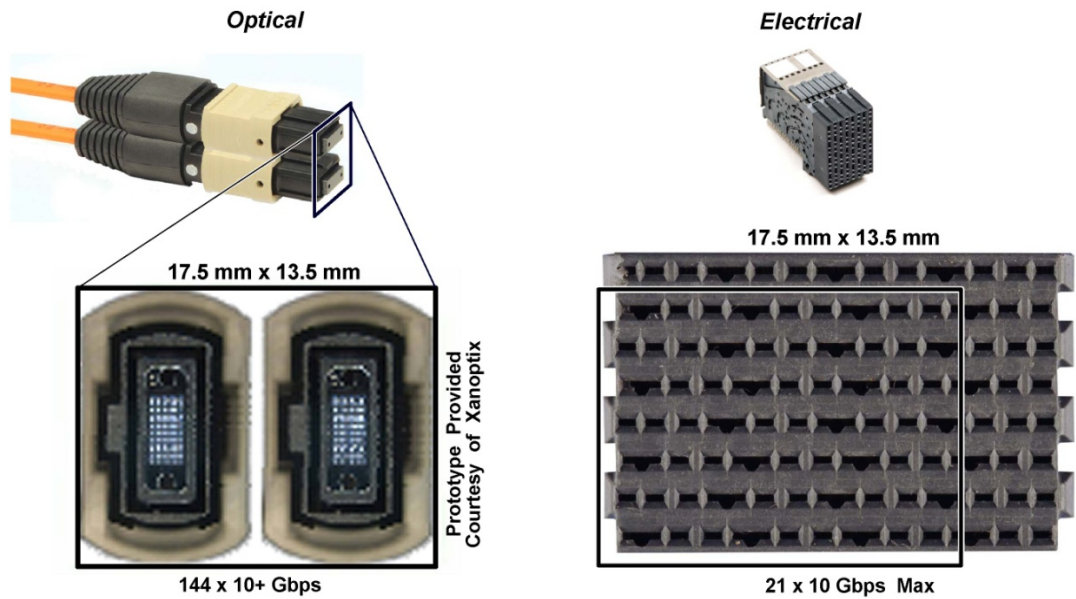

**Figure 4: Comparison of Multimode Optical vs. Electrical Interconnect Packaging Density (41341)**

The multimode example shown uses an MPO (Multiple-Fibre Push-On/Pull-off) connector with dual arrays of 6 x 12 fibers. These connectors can also be used with single-mode fibers. One example of this is a Luxtera transceiver using eight fibers with one wavelength per fiber [14]. The resulting interconnect density in this case is lower than the electrical connector, as shown in Table 1.

| Connector Type | Data Rate (Channels/Fiber x Fibers x Rate) (Gbps) | Cross-sectional Area (mm2) | Interconnect Density (Gbps/mm2) |
|---|---|---|---|
| Electrical | 1 x 21 x 10 | 17.5 x 13.5 | 0.889 |
| Multimode MPO | 1 x 144 x 10 | 17.5 x 13.5 | 6.095 |
| Single-Mode MPO (Luxtera) | 1 x 8 x 10 | 13.5 x 8.5 | 0.697 |
| Single-Mode MPO (Luxtera extended) | 40 x 8 x 10 | 13.5 x 8.5 | 27.887 |
| Single-Mode LC | 40 x 1 x 10 | 7.36 x 4.52 | 12.034 |

**Table 1: Interconnect Density For Different Connector Types.**

However, the interconnect density for single-mode fibers can be increased by using more than one wavelength per fiber. Potential examples of this are listed in Table 1. In the first case, the Luxtera example is extended to 40 wavelengths per fiber resulting in an interconnect density that is roughly twice that of the multimode case. The second case uses an LC connector which has only one fiber but has 40 wavelengths on that fiber. Increasing the number of wavelengths per fiber greatly increases the interconnect density but it also increases the Tx and Rx complexity. For a system design, the trade-off between this added complexity and the interconnect density would need to be studied to choose the optimal interconnect technology.

# Test Vehicle Descriptions

Two main types of optical all-to-all communication architectures in HPCs are "broadcast and select" and point-to-point. In the broadcast and select architecture, the data from every node is broadcast to all of the other nodes and each node then decides which data to receive and which data to ignore. In the point-to-point architecture, data from each node is sent to all of the other nodes through individual point-to-point links. We implemented both of these architectures in our optical test vehicles to model the all-to-all connections within an HPC. The purpose of these demonstration vehicles was to evaluate the strengths and weaknesses of each architecture and to compare each to all-electrical implementations. Note that in a complete system implementation, many (e.g. 10s to 100s to 1000s) of identical nodes would be integrated. In order to keep the level of integration to a practical level for this study, only 4-node systems were considered, and in some cases, the nodes were not identically constructed, such that a variety of optical components could be assessed.

Several different tests were done to measure the performance of each test vehicle. The primary measurements included link bit error rate testing from a given electrical input in each candidate system to the electrical output at the far end destination (with various optical components in between), and eye diagram measurements at the optical input to the receiver. The BER testing included determination of the overall link margin. In addition to this testing, each of the components was individually tested and characterized.

Bit error rate testing was done using a pattern generator to drive the optical transmitter with a pseudo-random bit sequence (PRBS) and an error detector at the receiver electrical output. The eye patterns were measured by driving the transmitters with a PRBS and measuring the optical signal at the input to the receiver using a sampling scope equipped with an optical input module and eye pattern analysis software. The link margins were tested by inserting a precision programmable attenuator into the optical path and measuring the bit error rate as a function of received optical power.

## Broadcast and Select Test Vehicle

Our first demonstration test vehicle was designed to evaluate the unique optical link capabilities of the broadcast and select architecture. This architecture puts multiple wavelengths onto one fiber which is then split multiple ways to broadcast the data to all of the nodes. The component that combines the different wavelengths onto one fiber and broadcasts the multiple wavelengths out to multiple nodes is known as a star coupler. Star couplers typically have n input fibers and n output fibers, where n typically ranges between 2 and 64. Any wavelength that is on any of the input fibers is distributed to all of the output fibers. Although an electrical analog to the star coupler was used in system designs over 25 years ago[18], an electrical equivalent to this signal distribution no longer exists for the high-speed signals prevalent in today's HPC designs.

Using single-mode optics allowed us to use dense wavelength division multiplexing (DWDM) in this test vehicle. With DWDM, a wavelength spacing of 100 GHz is possible which allows approximately up to 80 unique wavelength channels to be available within the optical C-band (1525-1565 nm). The DWDM technology was developed for optimizing waveguide bandwidth for long-haul telecommunications applications.  We use eight of those channels for our four-node demonstration.

The block diagram of our four-node broadcast and select test vehicle is shown in Figure 5:

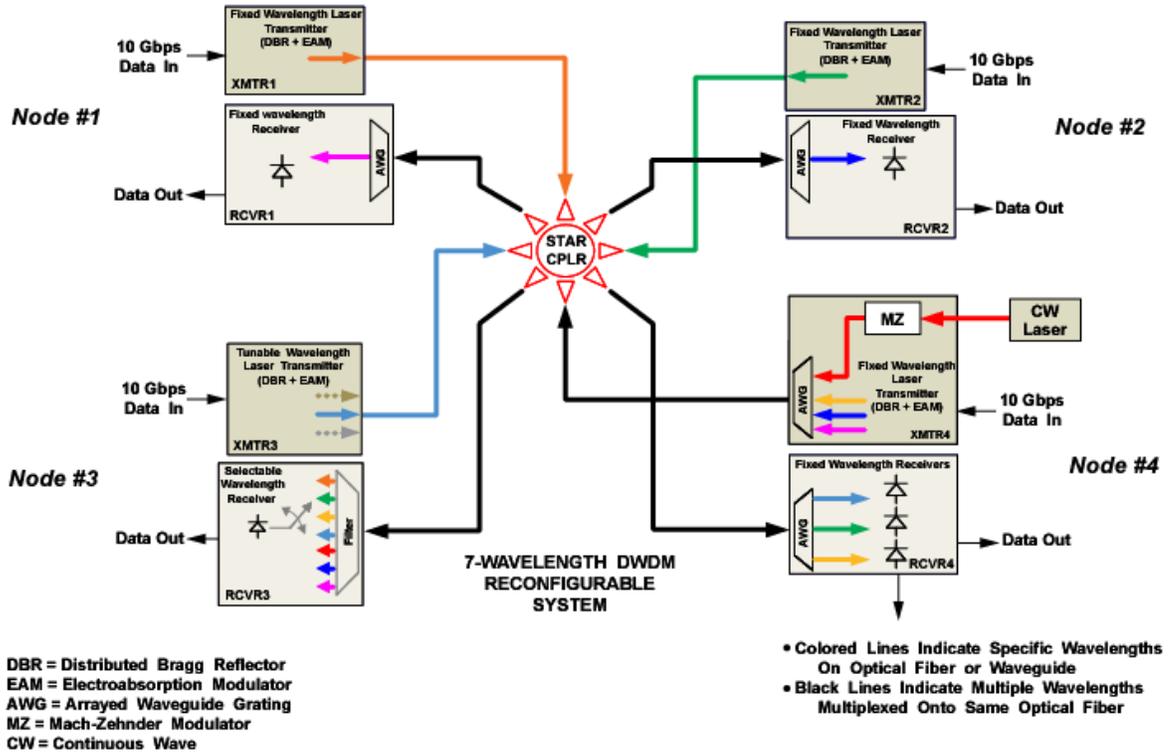

**Figure 5:  Schematic Diagram of Single-Mode DWDM Test Vehicle (41348)**

Each node consists of from one to four transmitters, and one to three photodetectors with an element for wavelength selection.  The wavelength selection is performed by an array waveguide grating (AWG).  The AWG is an n x m passive reciprocal component where n and m are generally from 1 to 16, although higher values are possible.  Light of a given wavelength entering on any of the n input fibers will be sent out the output fiber corresponding to that wavelength.  Being a reciprocal device, if light of the correct wavelength enters an output fiber, it will be transmitted onto all of the input fibers with a corresponding loss of 3 dB per factor of two in the number of ports (n) plus excess loss.

Two of the nodes (nodes 1-2) each have one transmitter which is tuned to a specific ITU channel.  The third has a single tunable transmitter which can be tuned to one of eight channels.  The fourth node has four transmitters, each tuned to a specific channel, which are combined onto a single output fiber using an AWG.  The channels at each node are listed in Table 2.

| Node | Transmitter | | Receiver | |
|---|---|---|---|---|
| | ITU Channel | Wavelength, nm | ITU Channel | Transmitter Node |
| **1** | 33 | 1550.92 | 30 | 4 |
| **2** | 34 | 1550.12 | 31 | 4 |
| **3** | 30-37 tunable | 1553.33–1547.72 | 30-37 selectable | Any |
| **4** | 30, 31, 32, 37 | 1553.33, 1552.52, 1551.72, 1547.72 | 32, 34, 35 | 4, 2, 3 |

**Table 2: ITU Channels for the Transmitters in Each Node.**

In Figure 5 the output fiber for transmitter is color coded to represent its specific wavelength. Seven standard wavelengths are thus transmitted on four fibers simultaneously to the inputs of the 4 x 4 star coupler.

In the star coupler, the different wavelengths are combined and broadcast on the star coupler's four output fibers simultaneously. They are then available on the photoreceiver's input fiber at each node. The star coupler's output fibers are color coded black in Figure 5 to represent multiple wavelengths being carried on those fibers.

Two of the nodes (nodes 1, 2) have an AWG acting as an optical filter at the receiver to select a specific wavelength (corresponding to a specific node transmitter). One node (node 4) has an AWG acting as an optical filter at its receiver which selects three wavelengths to receive. Node 4 also includes provisions for a Mach-Zehnder (MZ) modulator coupled to an external continuous wave (CW) variable wavelength laser source to allowing remote CW optical sourcing. The last node (node 3) has a wavelength selectable filter which is capable of selecting any one of the seven wavelengths that are available at that node. Table 2 lists the channels being received by each node and the corresponding transmitter for these channels.

## Multiple Wavelengths

As can be seen in Table 2, node 3 can transmit on any of the eight available channels that are not being used by the other transmitters. In Figure 6 below this capability is illustrated through multiple optical spectrum analyzer (OSA) scans of the output of the star coupler with the tunable transceiver set to unique ITU channels.

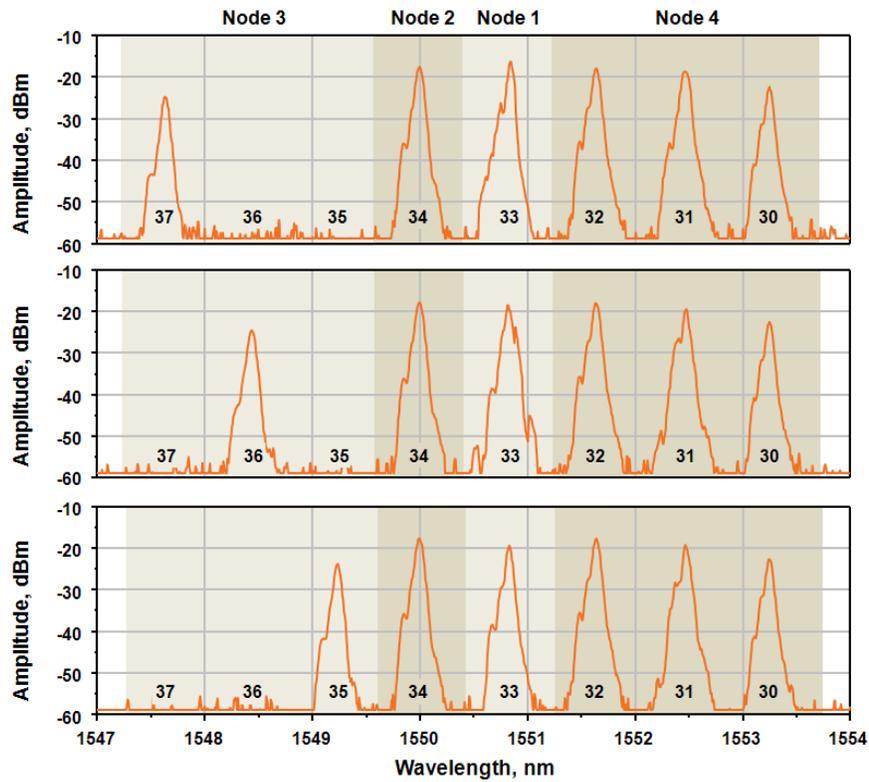

**Figure 6: Illustration of the ITU Wavelengths and Transmitter Wavelength as Displayed on the Optical Spectrum Analyzer (OSA) (42077)**

The entire broadcast and select test vehicle, along with the supporting lab characterization equipment is shown below in Figure 7:

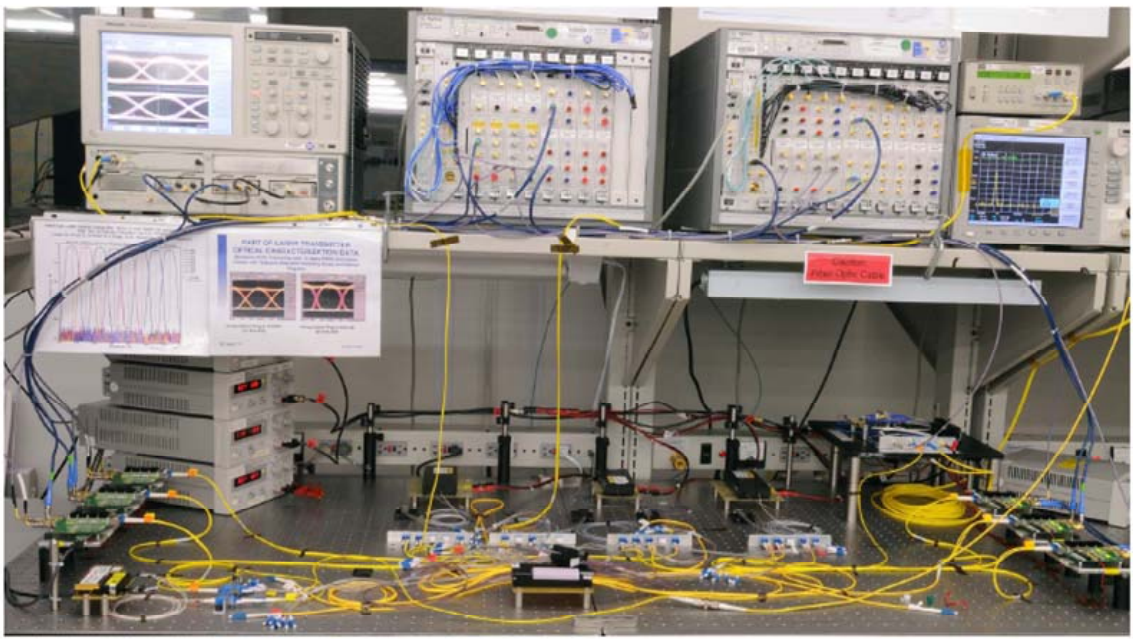

**Figure 7: Photograph of DWDM System Physical Layout (Benchtop) with the Star Coupler in the Center Foreground and Test Instrumentation (Shelf) (41653)**

### Photodetector

For the single-mode test vehicle, Avalanche Photo Detectors (APDs) were used at the optical inputs to the receivers. The increased input sensitivity (approximately -25 dB) for the APD's provided an overall link margin improvement of ~7dB when compared to similar links designed with P-doped, Intrinsic, N-doped (PIN) Photo diodes. The effects associated with the receiver decision are illustrated in greater detail in the section on link margins later in this report.

### Single-Mode Transceiver Power

Since the transceivers utilized for the DWDM test vehicle were essentially components targeted originally for long-haul telecommunication applications, their energy/bit characteristics were least attractive for the HPC demonstration. This higher power consumption is primarily due to the extra power associated with auxiliary circuits such as clock data recovery and the thermo-electric cooler (TEC) that precisely controls the transmitting source wavelength. A breakout of the transceiver power consumption and energy per bit transferred are shown below in Figure 8.

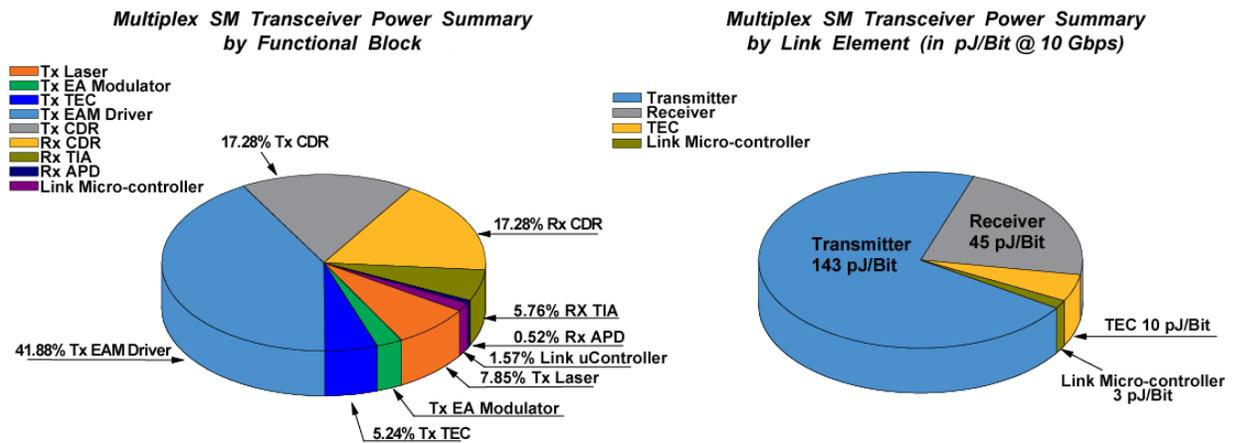

**Figure 8: Detailed Breakout of Power Consumption Distribution for the Multiplex DWDM Transceivers by Functional Block and Energy per Bit Transferred [pJ/bit] (41829)**

As can be seen from this illustration, approximately 34% of the power for the link is committed to the CDR circuits for transmitter and receiver. From a normalized power consumption perspective the transmitter alone accounts for 143 mW/Gbps (143 pJ/bit) at 10 Gbps. While the energy per bit transferred appears significantly outside of the range of consideration for HPC designs (10 pJ/bit) we have already seen signs of transmitters emerging at 40 Gbps. Assuming no significant additional power consumption, then the normalized power consumption for the single-mode transmitter could be reduced by a factor of four, down to roughly 36 pJ/bit and with further optimizations for the Tx and Rx CDRs, it may be possible to get nearer to the 10-20 pJ/bit range with the single-mode DWDM design.

## Point-to-Point Test Vehicles

Our next demonstration test vehicles were designed to evaluate the unique optical link capabilities of the point-to-point architecture. The lowest power and lowest cost options for this architecture are based on arrays of multimode vertical-cavity surface-emitting lasers (VCSELs) with a wavelength of 850 nm and arrays of photodetectors. This approach has gained acceptance for short communication links (< ~ 10 m) because of the lower power consumption and cost when compared to the single mode transceivers shown previously. The transmission medium of these links can be multimode optical fiber, either in standard 12-fiber ribbon cables or embedded in flexible material, such as Kapton. Another possible waveguide option is a polymer waveguide embedded within the printed circuit board for on-board chip-to-chip links [9]. Objectives for the test vehicles included exploring manufacturability of multimode optical interconnect solutions for HPCs, determine whether typical multimode links have sufficient operating margin for high BER applications, and evaluate opportunities for significant signal density improvement (vs. electrical links).

### Point-to-Point Test Vehicle #1: Fujitsu Optical Transceivers

The first point-to-point test vehicle has four nodes with a Fujitsu optical transceiver at each node and optical fiber embedded in a flex material to implement an "all-to-all" architecture. The Fujitsu transceivers each have four VCSEL transmitters and four receivers. The optical flex circuit interconnects the four nodes using a configuration known as a "perfect shuffle" interconnect structure.

A schematic representation of the overall multimode test vehicle is shown below in Figure 9.

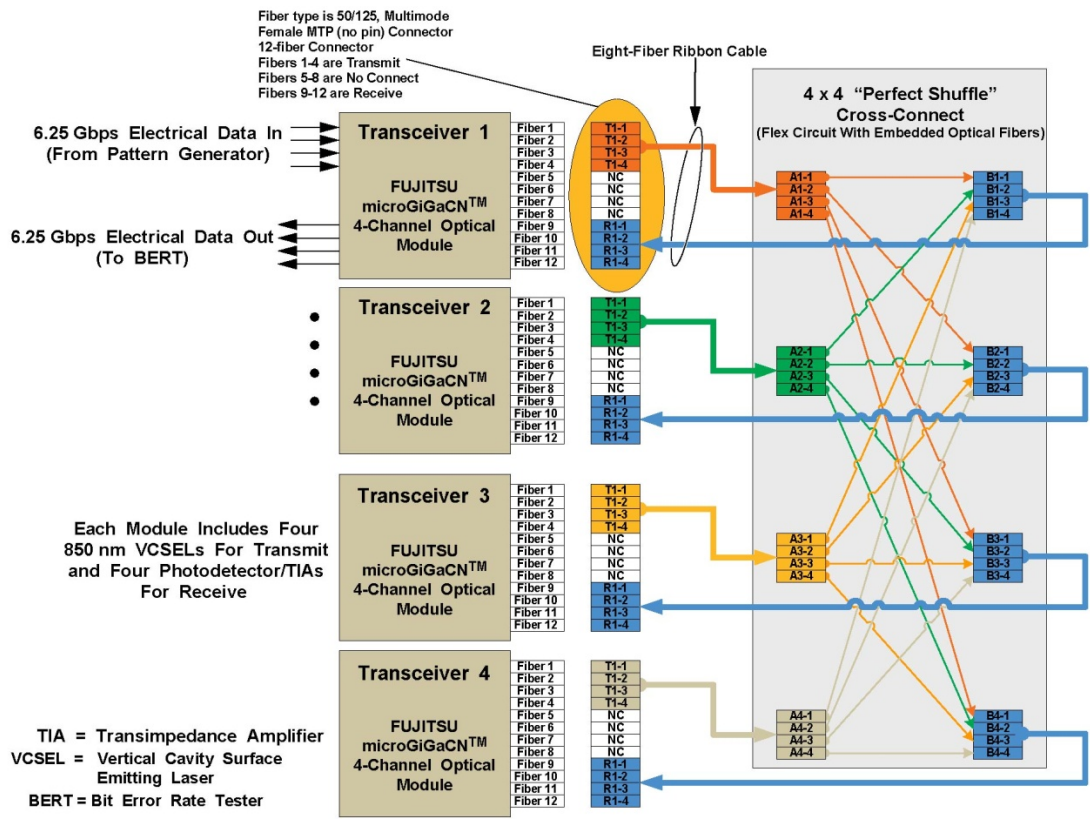

Figure 9: Dense Parallel Optics Demonstration Test Vehicle (41355)

And a photograph of the full multimode test vehicle is shown below in
Figure **10**.

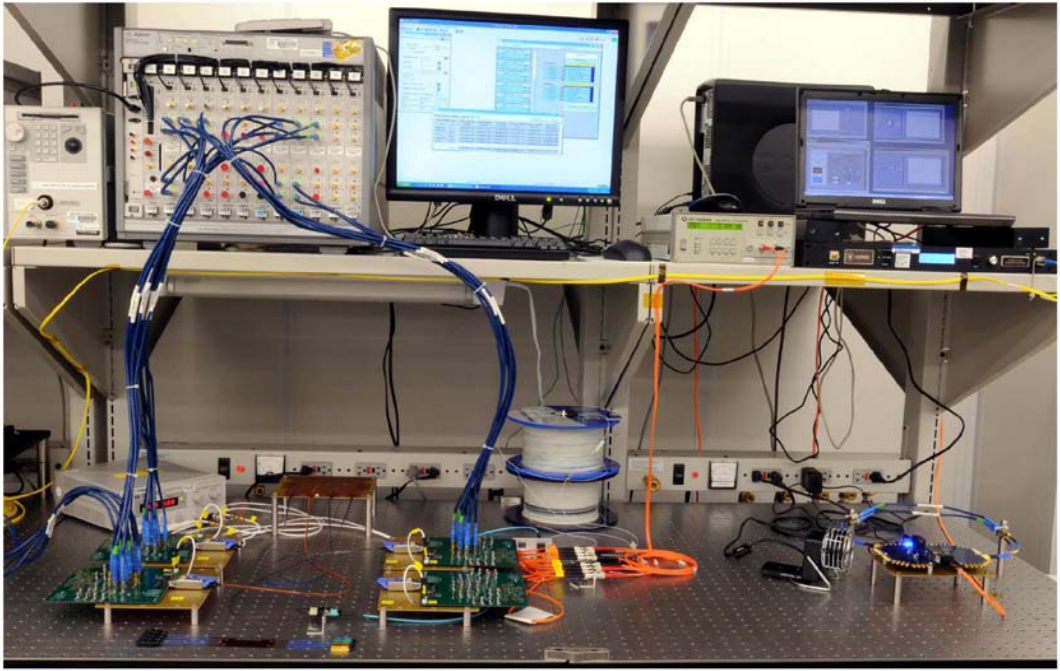

**Figure 10: Parallel Optics Demonstration Vehicle (41656)**

A close-up of the test card setup used to characterize the Fujitsu optical transceivers is shown below in Figure 11.

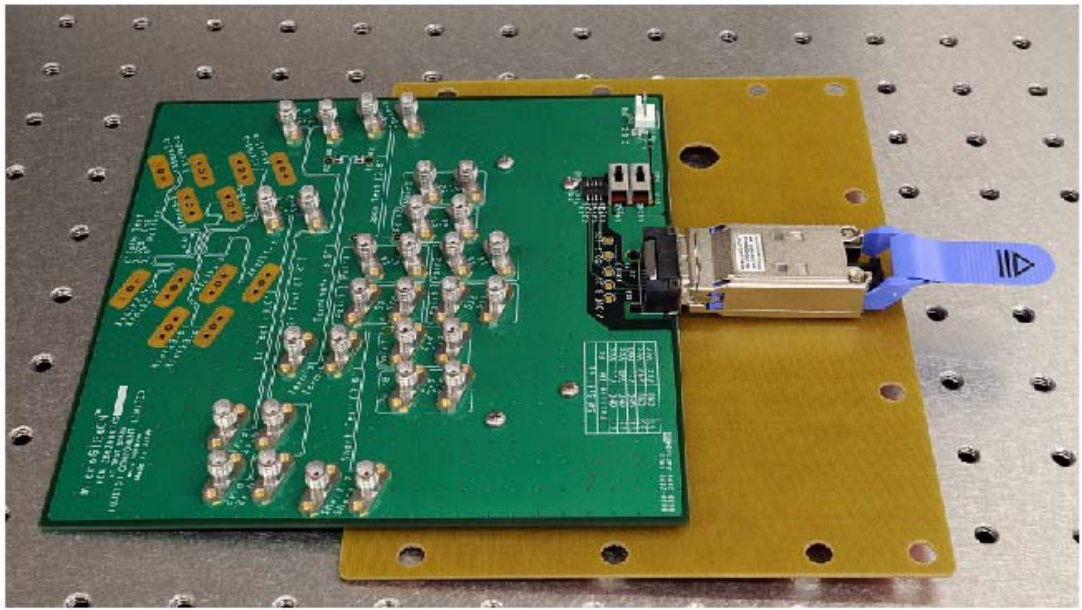

**Figure 11: Photograph of the Fujitsu Multimode Transceiver and Test Board Mounted on Support Plate (41804)**

## Point-to-Point Test Vehicle #2: Reflex Photonics Optical Engines

One of the trends in the application of optical interconnects to HPC systems is shorter and shorter links, i.e., not only are they becoming relevant for rack-to-rack and board-to-board interconnects, but as the data rates become even higher, intra-board optical links are being seriously considered. For such applications the optical engines need to be mountable anywhere on a printed circuit board (PCB) and possibly even directly mounted onto an application specific integrated circuit (ASIC) or field programmable gate array (FPGA) package. For these solutions, miniaturization of the electrical-to-optical (E/O) and optical-to-electrical (O/E) conversion "engines" becomes a necessity. An example of such technology is offered by Reflex Photonics.

### Reflex Photonics Optical Engine

Reflex Photonics optical engines are small and can be mounted on a PCB away from the edges of the board or package. The laser transmitter consists of an array of 12 VCSELs and associated driver circuitry, which converts 12 parallel electrical data inputs to 12 parallel optical output signals. The receiver module inputs 12 parallel optical signals and converts them into 12 parallel electrical signals through an array of 12 PIN photodiodes plus the associated receiver circuitry of transimpedance amplifier (TIA), electrical Serializer Deserializer (SerDes) and pre-emphasis channel compensation. Mayo-designed test PCBs were fabricated and sent to Reflex for mounting of their optical engines. One board has the laser transmitter and the second board the photoreceiver. The boards are connected with a 12-fiber ribbon cable. The resulting link has a potential capacity of 12 x 6.25 Gbps. The Mayo PCBs were designed to access four of these channels by providing four pairs of SMA connectors and corresponding differentially driven microstrip transmission lines that connect to the optical engines. Figure 12 shows the PCB with the optical engine encapsulated in a "glob top" seal.

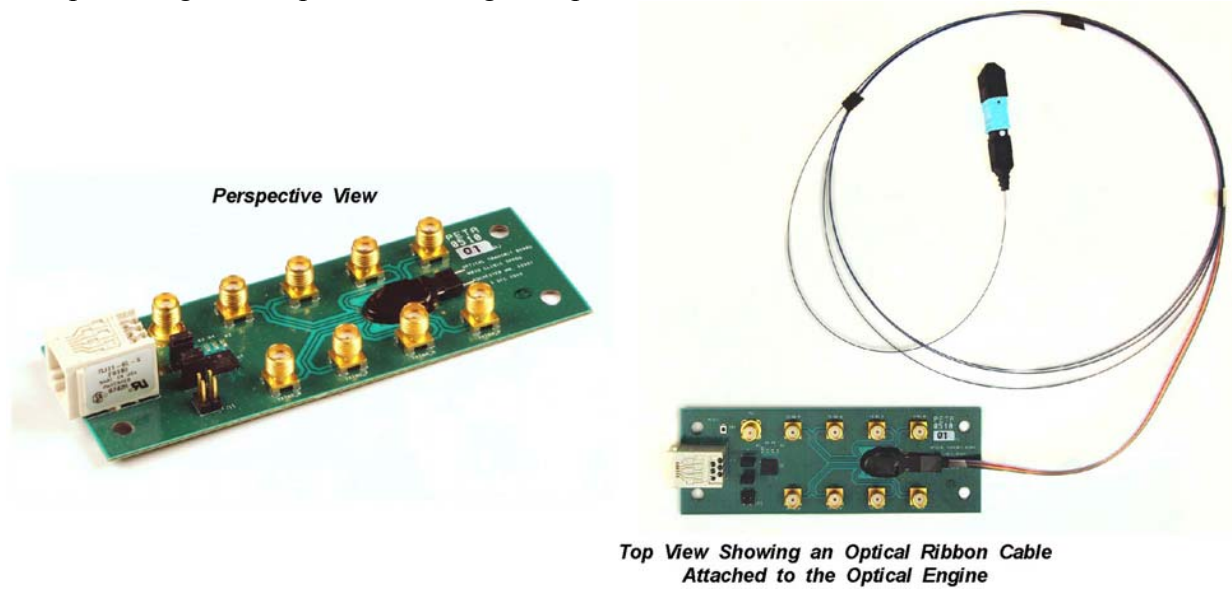

**Figure 12: Two Views of Reflex LightAble™ Optical Engine Mounted on Mayo-Designed PCB (41805)**

Figure 13 shows details of the internal structure of the optical engine and its interface with the fiber that is nominally sealed with the opaque glob top.

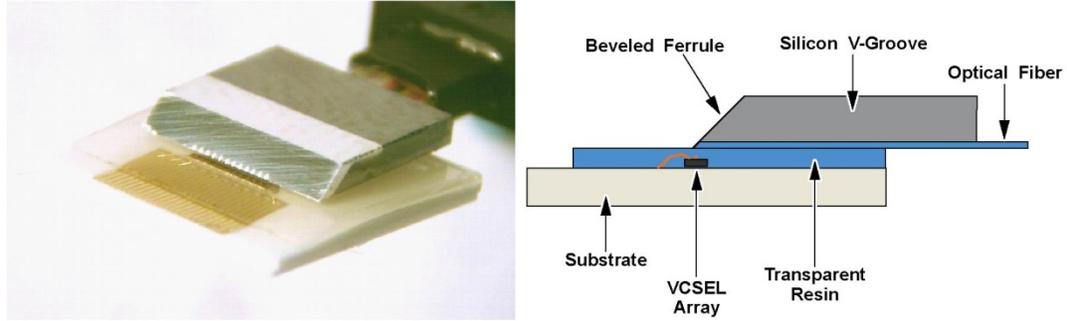

**Figure 13: Physical Realization of Reflex's Optical (VCSEL Transmitter) Engine (Left) and Corresponding Diagram of Optical Sub-Assembly (Right) Courtesy of Reflex Photonics (41830)**

The two Reflex test boards were used to demonstrate optical high-speed data transfer between them at 6 Gbps. Reflex is working on a 10 Gbps version that was not yet available for use on this project. Figure 14 shows a photograph of the Reflex optical link being driven by a pattern generator with the received 6 Gbps PRBS electrical signal from one channel displayed on the oscilloscope screen.

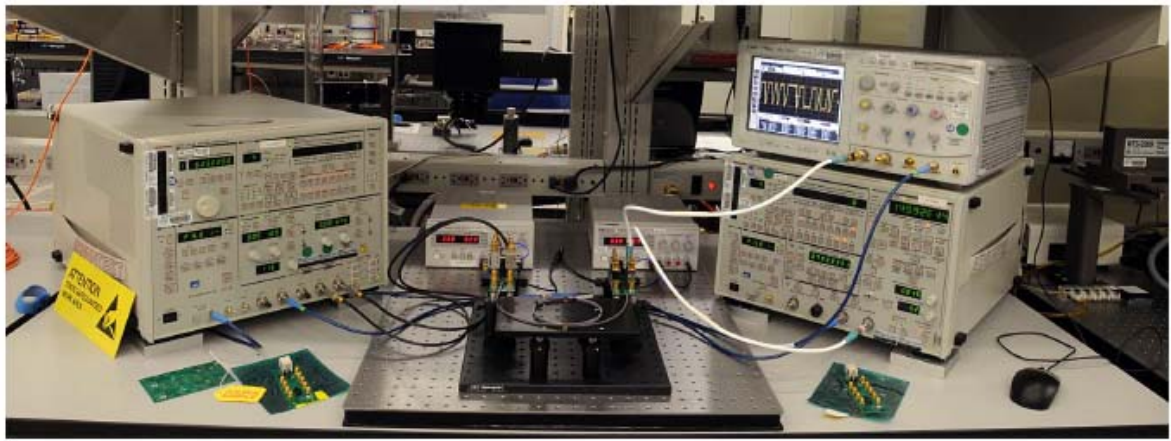

**Figure 14: Reflex Photonics Transmitter/Receiver Pair Under Test (the interconnecting optical ribbon cable is seen coiled up on the front of the support plate) (41806)**

# Test Vehicle Results
## Link Characterization Results and Data Transfer Efficiency

For the single-mode links, the average optical output power for the Tx without modulation was measured to be approximately 1.5 mW and the extinction ratio for these transmitters was measured at 16.2 dB. The total energy per bit transferred for the single-mode DWDM link exceeded 200 pJ/bit.

For the Fujitsu multimode test vehicle, the average optical output power from each transmitter is slightly less than 0.5 mW. The electrical input power to the transceiver is approximately 800 mW for four channels. At 6 Gbps this leads to a data transfer efficiency of 33.3 pJ/bit for a transmitter and receiver pair. For high speed operation VCSEL currents are typically not driven to near zero for NRZ modulation, and therefore, the extinction ratios are typically lower than they are for the DWDM transmitters. For the Fujitsu transmitter the extinction ratio is approximately 6 dB.

For the Reflex Photonics multimode test vehicle, the average optical output power with modulation measured 0.4 mW, and the extinction ratio is also approximately 6 dB. Power consumption for the 12 transmitters is approximately 1 W, and for the 12 receivers is 2.1 W. The resultant energy per bit transferred is 44 pJ/bit.

## Additional Characterization Results from the Test Vehicles
### Point-to-Point Links Within the Multimode Optical Flex Circuit

The all-to-all optical signal distribution for the multimode dense parallel optics demonstration was handled through the use of point-to-point fiber optic waveguides embedded within an optical "perfect shuffle" interconnect. The perfect shuffle structure is composed of multiple channels of multimode fiber optics overlaid and sandwiched between two laminated layers of Kapton polymer material for protection. The perfect shuffle element is shown below in Figure 15.

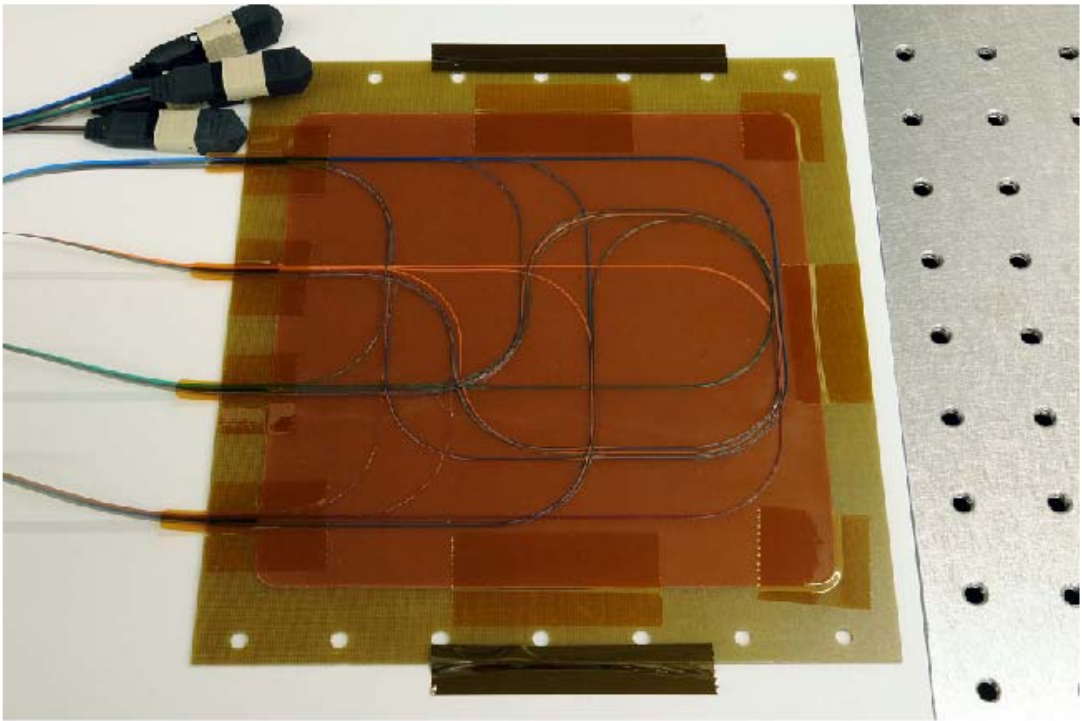

Figure 15: Photograph of the 4 x 4 "Perfect Shuffle" Optical Cross-Connect Network (41803)

The benefits of this approach for multimode optical signal distribution is that it is capable of accommodating the higher B*d characteristics associated with multimode fiber versus the B*d associated with PCB-embedded polymer optical channels. The former still provides channel isolation, while facilitating multiple direct waveguide crossovers within a total structure thickness of approximately 15 mm. The fibers within the flex circuit are brought out to four Multifiber Termination Push-on (MTP) optical connectors. Each connector utilizes 8 of the 12 available channels to support two sets of four simplex optical channels. The connectors facilitate direct connection to the Fujitsu multimode transceivers.

Since the time that this effort was conducted, we have also identified sources that are capable of fabricating multimode star couplers. We originally learned of the star coupler as a potential broadcast element for an HPC application courtesy of the Lockheed Martin optical engineering team[7]. We later identified a multimode version of the star coupler [19]. In principle it might be feasible to construct a four channel CWDM multimode optics test vehicle, although based on the results from multimode link margin testing, the potential system scalability would be limited to four compute nodes

## Comparison of the Optical Transmitter Output Eye Patterns

The optical output characteristics for all three test vehicles were characterized and the resultant eye patterns are shown below in Figure 16.

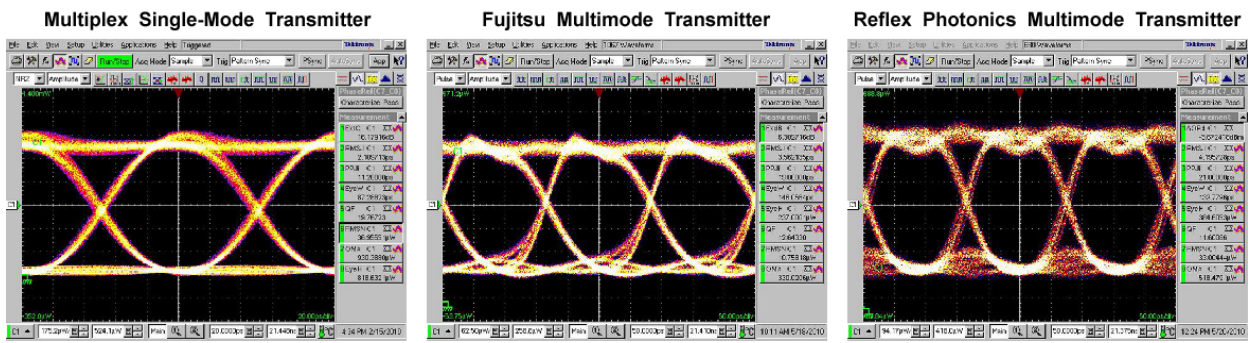

Figure 16: Comparison of Optical Eye Patterns for the Three Optical Test Vehicles (42068)

Even though it is transmitting a higher data rate than the other two optical solutions, the eye pattern for the DWDM transmitter shown on the left clearly shows much smaller variations in signal amplitude and timing jitter. The sharp DWDM eye pattern is also attributable to the high extinction ration achieved through the use of an electro-absorption modulator (EAM), integrated into the Multiplex Transceiver package.

## Link Margin Testing
### Link Margin Testing on the Multimode Links

Link margin testing for the multimode test vehicle was limited to testing a specific, yet representative, point-to-point channel from one of the sixteen independent channels available in the system (there are a total of four transceivers and each transceiver incorporates four transmitters and four receivers that are interconnected through independent point-to-point links via the optical shuffle cross-connect). Prior to testing a specific channel for link margin, the overall system was checked for nominal performance using the ParBERT system. Margin testing was conducted with the adjacent channels operating to include the effects of cross-coupling. A break-out optical cable assembly was used to separate out a nominal optical channel for direct optical eye pattern characterization. The data rate was set to 5 Gbps, and the nominal output modulation amplitude (OMA) was measured to be -6.7 dBm (215 µW) prior to introduction of the attenuator into the link. Link BERs versus link attenuation was then characterized and the results plotted. The test setup used for both link types are shown below in Figure 17, which also shows the link margin test results for the Fujitsu multimode link in the right-most side of the plot.

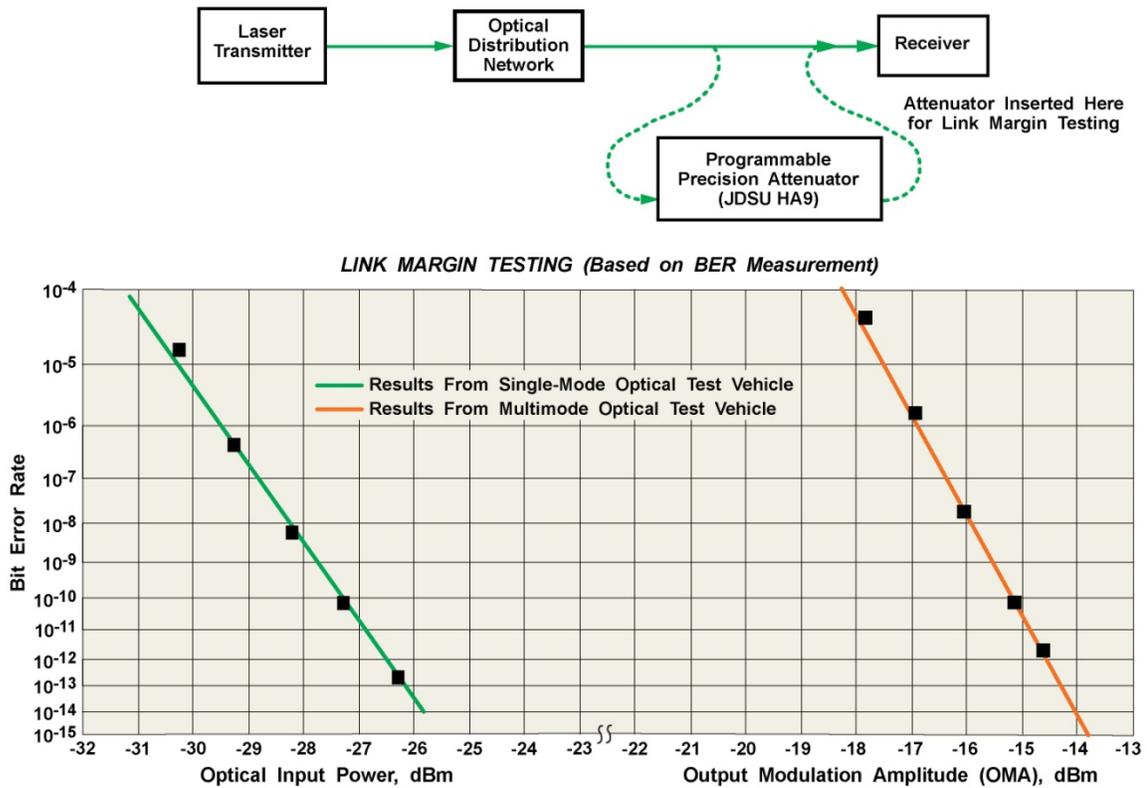

Figure 17: BER Test Results For The DWDM and Multimode Test Vehicles (42053)

The margin BER test results show that the OMA value at the $10^{-12}$ error rate is approximately -14.5 dBm.  Using the value for the nominal OMA of -6.7 dBm, the link margin is calculated to be approximately = -6.7 – (-14.5) = 7.8 dB.  While this is considerably smaller than the corresponding values for the DWDM system, the link margin is quite sufficient for the short reaches within an HPC system.  Operating link margins for the Reflex Photonics Test Vehicle were also measured to be equivalent to the Fujitsu links within the range of repeatability error.

## Link Margin Testing on the DWDM Optical Link
A test of the link margin for specific links incorporated in the broadcast-and-select DWDM system was also conducted using the Advantest pattern generator and error detector instruments and the Agilent ParBert system.  In these tests other channels not being driven by the pattern generator are transmitting independent data streams driven simultaneously by the ParBERT system, so that link margin for a specific link is measured under more realistic conditions.
The link margin result for the DWDM link is shown in the left-most graph in Figure 17.  The power measured at the input to the photoreceiver (prior to introducing the attenuator) was measured to be -8.4 dBm.  From the BER graph a bit error rate of $10^{-12}$ occurs for an optical input power of about -26.5 dBm.  Consequently the link margin for this specific link is calculated to be: -8.4 - (-26.5) = <u>18.1</u> dB.   These margin results confirm the effects of the improved input sensitivity associated with the APD receiver.

### Scalability of the Broadcast-and-Select Architecture to 32 Nodes
The significantly larger link operating margins of the DWDM test vehicle allowed us to demonstrate scalability of this vehicle up to the equivalent of a 32 node compute system by substituting a 32 x 32 star coupler prototype for the 4 x 4 unit.  The prototype unit was provided courtesy of the Lockheed Martin Corporation, Mission Systems and Sensors Division [16].  In this case, only four of the 32 inputs and four of the 32 outputs of the star coupler were utilized.

For this test case, the link margin = -20.7 -(-25.5) = <u>4.8 dB</u> for the DWDM system when the 32 x 32 star coupler is used as the broadcast element.  Note that this is consistent with the previous data using the nominal 4 x 4 star coupler with a link margin of 18.1 dB.  The difference in splitting loss between the 4 x 4 versus the 32 x 32 coupler alone accounts for 9 dB reduction in link margin (6 dB for the 4 x 4 and 15 dB for the 32 x 32 coupler).  In addition, the excess loss of the 32 x 32 coupler versus the 4 x 4 coupler accounts for at least another 4 dB in link margin reduction, yielding a total of 13 dB link margin reduction from the 18 dB value, or approximately 5 dB.  Given the variation in insertion loss for the optical connections alone, the agreement in values between expected (as extrapolated from the nominal system using the 4 x 4 coupler) and measured results for the 32 x 32 star coupler case is excellent.  In theory, it might be possible to accommodate the additional 3 dB splitting loss associated with scaling to a 64 X 64 star coupler, however we view the minimal 1.8 dB of remaining link margin as insufficient for consistent operation of the DWDM links at BERs of 1e-12, and therefore, that additional amplification would be required for such a system to be considered.

## Summary and Comparison of Results
During the course of this effort we assembled a table summarizing highlights of link power, energy per bit transferred, data rates and reach distance for some of the various examples of electrical, multimode optical and single-mode optical links encountered.  The results are shown below in Figure 18.

| Vendor or Reference | Wavelength [nm] | Channels x Bit Rate n x [Gbps] | Power per Transceiver [mW] | Energy/Bit [pJ/bit] | Approximate Reach [m] | Bandwidth-Distance Product - B*d Each Channel [Gbps-cm] |
|---|---|---|---|---|---|---|
| **Optical Links** | | | | | | |
| Optical Link Supplier #1 | 850 (VCSEL-based) | 4 x 10 and lower rates | 1500 | 38 | 100 on 4-lanes of MM fiber | 10,000 |
| Optical Link Supplier #2 | 850 (VCSEL-based) | 4 x 5 and lower rates | 1050 | 53 | 100 on 4-lanes of MM fiber | 5,000 |
| Optical Link Supplier #3 | 850 (VCSEL-based) | 4 x 10 and lower rates | 750 | 19 | 100 on 4-lanes of MM fiber | 10,000 |
| Optical Link Supplier #4 | 1550 (DFB Laser-based) | 4 x 10 and lower rates | 2200 | 55 | 300 on 4-lanes of SM fiber | 30,000 |
| Optical link research result (2007) from IBM (Terabus) 130 nm CMOS for drive electronics [9] | 985 (VCSEL-based) | 16 x 10 (up to 16 x 15) and lower rates | 2160 (at 10 Gbps) | 13.5 | ≤1 on 16 lanes of embedded optical waveguides in FR4 (Optoboard) | 1,000 |
| **Electrical Links** | | | | | | |
| All-electrical research result (2007) from Rambus, Inc. in 90 nm CMOS [10] | na | 6.25 only | 14 | 2.2 | 0.8 (200 μm microstrip on FR4 medium) | 500 |
| All-electrical research result (2008) from Intel in 65 nm CMOS [11] | na | 1 x 15 and lower rates (Power Managed) | 75 | 5 | 0.20 microstrip on FR4 medium | 300 |
| All-electrical research result (2009) from Rambus, Inc. in 65 nm CMOS [12] | | 512 x 16 | 172,000 | 21 (private communication) | 0.08 microstrip on FR4 medium | 128 |

DFB = Distributed Feedback
VCSEL = Vertical Cavity Surface Emitting Laser

**Figure 18: Summary of Normalized Power For Optical and Electrical Interconnect Solutions (40342)**

By further examining these results, we see that B*d for electrical PCB constructs is the lowest, and was also found to be very dependent upon characteristics such as the PCB materials utilized, care in the PCB fabrication process, and the relative performance (and hence cost) associated with the interconnect components and methods utilized. Polymer-based optical waveguides are seen to have improved B*d when compared to electrical constructs, although the limited improvement seen makes the high costs associated with" hybrid" optical/electrical PCB constructs questionable at this point in time/. Until the transmission characteristics of polymer-based optical waveguides are significantly improved, Multimode and Single-mode fiber optical constructs will most likely be the design solution chosen for longer transmission lengths and higher bandwidth link requirements within HPC designs. Also shown within this chart is that electrical links still clearly have a significant power/efficiency advantage when compared to optical links.

Based upon the metrics described earlier within this paper, we have examined and compared the results seen across this effort, to results seen from numerous other efforts conducted within our organization regarding electrical link characterization [17]. We have assigned a relative 1-5 (1=lowest, 5= highest) ranking to compare the four possible link types described, and the results are summarized below in Table 1:

| Metric | Electrical Links | Single-Mode Optical Links | Fiber-Based Multimode Optical Links | Polymer-Based Multimode Optical Links |
|---|---|---|---|---|
| B*d | 1 | 5 | 4 | 2 |
| Power | 4-5 | 1 | 3 | 3 |
| Cost | 4-5 | 1 | 3 | 2 |
| Density | 1-2 | 2 Current<br>5 Silicon Research | 4 | 5 |

Table 3: Comparison Summary of Four Types of Interconnect Solutions (42074)

# Conclusions

Two fundamentally different types of optical data communication demonstration systems were assembled and tested using commercially available, and in some cases experimental, state-of-the-art components. One is based on DWDM single-mode optics with a "broadcast-and-select" network using edge-emitting lasers while the other two are based on VCSEL transmitters and "parallel optics" utilizing strictly point-to-point optical communication links. Both types of approaches are potentially applicable to High Performance Computing (HPC) systems. For reference, we also compare research results reported [9] regarding a third optical approach utilizing polymer-based multimode waveguides. All three link types were then compared to traditional electrical based high-speed serial links.

One of the metrics that clearly favored the optical links over electrical links was the bandwidth-distance product. It is our opinion that this may be the most influential metric associated with seeing a crossover occur from electrical to optical signaling, as more systems with specific requirements that exceed the current range of B*d for electrical interconnects are forced to seek an optical interconnect solution as their only choice.

A second advantage for optics was seen in the comparison of the packaging density, where a multimode optical interconnect showed more than a 6X advantage over one high-speed electrical counterpart. When composed from components optimized for telecommunication applications, single-mode optical fiber-based constructs do not exhibit favorable metrics in terms of packaging density, however there are many corporate laboratory research efforts currently under way that are targeting optical link solutions embedded within silicon substrates that may be capable of demonstrating significantly higher signal densities in the future.

Key advantages still favoring electrical links were found with respect to both cost (in $/Gbps) and power consumption, although multimode optical links were also found to be making improvements in both areas.

For link application within HPC systems we view multimode optics as having greater near-term viability than the single-mode DWDM system primarily because of lower cost, higher efficiency, and better packaging density, however electrical links will retain the dominant position in HPC interconnects until lower cost, lower power optical solutions within this environment are demonstrated.

# Acknowledgements
The authors would like to acknowledge several people who contributed to the paper in important ways. In particular, we would like to thank Gregory Whaley, and the optical engineering team at Lockheed Martin, for early access to custom optical components as well as extended technical consultations. We would also like to thank Ming-Lai of Multiplex for timely delivery of DWDM transceivers; Brian Lindroth (of High Technology Sales) and Fujitsu for providing access to Fujitsu multimode evaluation boards. We also thank Kerry Holm for his timely support in designing the Reflex Photonics test modules, Vic Gammell, Chuck Burfield, and Mike Lorsung of Mayo's SPPDG laboratory for continued and enthusiastic technical support, and our warmest thank you to the Mayo graphic art team: Steve Richardson, Elaine Doherty, Terri Funk and Deanna Jensen for all their long hours of support for throughout the program.

# References
[1] Astfalk, G., "Why optical data communications and why now?" Appl Phys A (2009) 95: 933–940 DOI 10.1007/s00339-009-5115-4
[2] Krishnamoorthy, A. V., et al., "Computer Systems Based on Silicon Photonics Interconnects," Proceedings of the IEEE, vol. 97, No 7, pp. 1337-1361, July 2009
[3] Tan, M., et al., "A high-speed optical multi-drop bus for computer interconnections" Appl Phys A (2009) 95: 945–953 DOI 10.1007/s00339-009-5162-x
[4] Mussman, H.: Using Optical Links to Interconnect Digital Equipment. Panel Session, International Solid State Circuits Conference (ISSCC) 85, February 14, 1985
[5] Lemoff, B., et al; MAUI: Enabling Fiber-to-the-Processor With Parallel Multiwavelength Optical Interconnects, IEEE Journal of Lightwave Technology, Vol.22, No. 9, pp 2043-2054, Sept., 2004.
[6] Active Optical Cables Markets and Applications reproduced with permission from Dr. H Pan; http://www.nefc.com/wp-content/uploads/nefc_20100518.pdf
[7] Horne, M.A., Schantz, H.J., Newcomer, S.O., Whaley, G.J., "Air Force highly integrated photonics (HIP) program", Proc. SPIE, Vol.6243, 624301 (2006).
[8] Trezza, J., et al.: Parallel Optical Interconnects for Enterprise Class Server Clusters: Needs and Technology Solutions. IEEE Optical Communications, pp. s36 – s42, February 2003.
[9] Doany, F. E., et al.: Terabus: A 160-Gb/s Bidirectional Board-Level Optical Data Bus. Conference Proceedings of the 20th Annual Meeting of the IEEE Lasers and Electro-Optics Society, pp. 545-546 (October) 2007.
[10] Poulton, J., et al,: A 14 mW 6.25-Gb/s Transceiver in 90-nm CMOS. IEEE Journal of Solid State Circuits, 42(12):2745-2757 (December) 2007.
[11] Balamurugan, G., et al.: A Scalable 5-15 Gbps, 14-75 mW Low-Power I/O Transceiver in 65 nm CMOS. IEEE Journal of Solid State Circuits, 43(4):1010-1019 (April) 2008.
[12] Beyene, W.T., et al.: The Design and Signal Integrity Analysis of a TB/sec Memory System. DesignCon 2009 Conference (February) 2009.
[13] Fisher, J.: Optoelectronic Substrates: Will They Happen? Journal of Circuits Assembly (November) 2007; Co, 1992.
[14] Luxtera OptoPHY LUX6004 40 Gigabit Optical Transceiver: http://www.luxtera.com/optophy-lux6004.html.
[15] Pepeljugoski, P., et al.: Comparison of Bandwidth Limits for On-Card Electrical and Optical Interconnects for 100 Gb/s and Beyond. Proceedings of SPIE, Optical Interconnect Technologies II: Joint Session With Conference, 6897:68970I(7 pages), January 21, 2008, San Jose, CA.
[16] Whaley, G, et al; Air Force Highly Integrated Photonics program: Development and demonstration of an optically transparent fiber optic network for avionics applications," SPIE Vol. 7700, 77000A · © 2010 SPIE CCC code: 0277-786X/10/$18
[17] McCoy, B., et al,: PWB Manufacturing Variability Effects on High Speed SerDes Links: Statistical Insights from Thousands of 4-Port Parameter Measurements. DesignCon 2010 Conference (February) 2010.
[18] Wikipedia definition of "Star Coupler": http://en.wikipedia.org/wiki/Star_coupler.
[19] Optical interLinks website: http://www.opticalinterlinks.com